\documentstyle[floats,prl,aps,preprint]{revtex}
\begin{document}
\tightenlines
\preprint{UAB-FT-441}

%%%%%%%%%%%%%%%%%%%%%%%%%%%%%%%%%%%%%%%%%%%%%%%%%%%%%%%%%%%%%%%%%%%%%%
%
%  Uncomment following four lines and one below for 2 column format
%  and figure insertions.
%

%\twocolumn[\hsize\textwidth\columnwidth\hsize\csname
%@twocolumnfalse\endcsname
%%%%%%%%%%%%%%%%%%%%%%%%%%%%%%%%%%%%%%%%%%%%%%%%%%%%%%%%%%%%%%%%%%%%%%
\title{Smooth ``creation'' of an open universe in five dimensions}
\author{Jaume Garriga \footnote{garriga@ifae.es}}

\address{IFAE, Departament de Fisica,\\
         Universitat Autonoma de Barcelona,
         08193 Bellaterra, Spain.}
\date{\today}
\maketitle
\begin{abstract}
We present a non-singular instanton describing the creation of an
open universe with a compactified extra dimension. 
The four dimensional section of this solution is a singular instanton
of the type introduced by Hawking and Turok. The
``singularity'' is viewed in five dimensions as a smooth bubble of
``nothing'' which eats up a portion of spacetime as it expands. 
Flat space with a compact extra dimension is shown to be gravitationally 
metastable, but sufficiently long lived if the size of the extra 
dimension is large compared with the Planck length.
\end{abstract}

%%%%%%% Comment the next line before submission
%\vskip2pc]
%%%%%%%%%%%%%%%%%%%%%%%%%%%%%%%%%%%%%%%%%%%%%%%%%%%%%%%%%%%%%%%%%%%%%%

Hawking and Turok have recently proposed that
an open universe may be created from nothing \cite{HT}
This is a very interesting possibility since it would 
lead to open inflation without requiring a
special form of the inflaton potential. This is an 
advantage over existing models, where part of the inflaton potential
has to be suitable for tunneling in one region and suitable
for slow roll in another. The price to pay, however, is
that the instanton describing this process has a singular boundary
(which is time-like in the Lorentzian region).
Nevertheless, the Euclidean action of the instanton is 
integrable, and the boundary term is finite. Therefore, Hawking 
and Turok were able to proceed formally, and used their instanton to 
assign probabilities to different open universes.

Naturally, the use of singular instantons has raised some concern
\cite{all,alex,unruh}.
For instance, it has been argued \cite{unruh} that information may 
flow in and out from the singularity, which would mean that
predictions cannot be made in that spacetime. On closer examination, 
however, it turns out  that the singularity behaves as a reflecting 
boundary for scalar and tensor cosmological perturbations \cite{ga98,GMTS}. 
Hence, the cauchy problem 
seems to be well posed and the model is well suited for the quantization of 
small perturbations and for comparison with observations.

Even so, it is clear that singular instantons cannot be
used without further justification. Indeed, Vilenkin has 
shown \cite{alex} that an 
instanton with the same singularity as Hawking and Turok's would lead to 
the immediate decay of flat space, in contradiction with observations.
Hence, there is some question as to whether singular instantons, 
even if integrable, can be used to describe the creation of an open universe. 

On the other hand, Vilenkin's argument provides an 
interesting clue towards fixing the problem. The unsuppressed decay of flat 
spacetime is due to the fact that singular instantons can have an arbitrarily 
small size, so that their action is as small as desired. It is therefore 
possible that if the model has a length scale below which Physics is 
different, the instability can be traded for metastability with a low decay 
rate. 

In this paper, we present a non-singular five dimensional model 
where some of the nice properties of the singularity in four dimensions can be 
understood and where the unsuppressed decay of flat space is avoided. 
As we shall see, flat space is metastable, but its decay rate will be 
exponentially small provided that the size of the extra dimension is much 
larger than the Plack length.

\section*{Metastable flat spacetime}

Let us show that flat space with an extra dimension 
is gravitationally metastable. It decays through the nucleation
of bubbles of ``nothing'' which eat up spacetime as they expand.

The five-dimensional action for pure gravity is given by
\begin{equation}
S_E= -{1\over 16\pi G_5} \int \sqrt{\tilde g} \tilde{\cal R} d^5x
     -{1\over 8\pi G_5} \int \sqrt{\tilde\gamma} \tilde K d^4 \xi,
\label{action5}
\end{equation}
where $G_5$ is the five-dimensional gravitational coupling,
$\tilde{\cal R}$ is the Ricci scalar and the last
term is the integral over the boundary of the trace of the
extrinsic curvature $\tilde K$. The tilde distinguishes five-dimensional
quantities from their four-dimensional counterparts, which we shall encounter
in the next section. 

Taking an $O(4) \times U(1)$ symmetric ansatz for the metric
\begin{equation}
d\tilde s^2= d\tau^2 + R^2(\tau) dS^{(3)} + r^2(\tau) dy^2,
\label{fivemetric}
\end{equation}
where $dS^{(3)}=(d\psi^2 + \sin^2\psi d\Omega_2^2)$
is the metric on the three-sphere and $y$ is the coordinate in the
fifth compact dimension,
the equations of motion reduce to \cite{fre82}
$\ddot X = 2 k$ and $r \ddot R - \dot r \dot R =0$,
where $X\equiv R^2$, $k=1$ is the spatial curvature of the 3-sphere and dots 
indicate derivative with respect to $\tau$. The first equation
indicates that $R^2$ is quadratic in $\tau$. The second tells us that 
$r$ is proportional to $\dot R$. The constant of proportionality is
unimportant, since it can be reabsorbed in a redefinition of $y$. 
Thus, the general solution is given by 
\begin{equation}
d\tilde s^2= d\tau^2 + (\tau^2+A^2) dS^{(3)} + 
\left({\tau^2 \over \tau^2+A^2}\right) dy^2.
\label{instanton}
\end{equation}
In what follows, we shall take $A^2>0$. \footnote{A duality transformation
takes $A^2 \to -A^2$.}

The instanton (\ref{instanton}) is 
perfectly regular.  At $\tau=0$ there is a ``polar'' coordinate singularity
in the $(\tau, y)$ plane, but the manifold is smooth and has no conical 
singularity if we take coordinate range in the fifth dimension as 
$0 \leq y < 2\pi A$.
The size of the extra dimension is zero at 
$\tau=0$ and goes to the constant value $A$ at large distances 
$\tau \to \infty$.  Thus the instanton
can be viewed as the direct product of a ``cigar'' times a three-sphere.
The size of the three-sphere tends to a constant $A$ at
$\tau=0$, and grows linearly with $\tau$ at large distances, like
it would in flat space. Thus our instanton is asymptotically flat.

Our solution (\ref{instanton}) is analogous to Coleman and De Luccia's
instanton \cite{coleman} describing the nucleation of a ``true vacuum'' bubble.
The important difference is that here there is no ``true vacuum'' to speak of. 
The interior of the 3-sphere of radius $A$ at $\tau=0$ contains no
spacetime: it is a bubble of ``nothing''.
The evolution of the bubble after nucleation is given by the analytic 
continuation of (\ref{instanton}) to the Lorentzian section. This is 
obtained by complexifying the angular coordinate 
$\psi \to (\pi/2)-i \hat\psi$,
where $\hat\psi$ is real. With this, the 3-spheres become 2+1 dimensional
time-like hyperboloids. The bubble grows 
with constant proper acceleration $A^{-1}$, eating up spacetime as it 
expands.

The nucleation rate can be estimated as
\cite{coleman}
\begin{equation}
\label{rate}
\Gamma\sim A^{-4} B^2 e^{-B},
\end{equation}
where $B= S_E-S_E^{flat}$ is the difference between the action of our
instanton minus the action of flat space.
Our instanton is a vacuum solution, so only the boundary term at infinity
($\tau\to \infty$) contributes to the action. (Clearly, there is no boundary
term at $\tau=0$, since the fifth dimension smoothly closes the manifold 
there.) This term can be expressed as the normal derivative of the 
volume of the boundary,
\begin{equation}
S_E= {-1\over 8\pi G_5} \int \partial_\tau\sqrt{\gamma} d^3 S^{(3)} dy,
\label{boundary}
\end{equation}
where $\sqrt{\gamma}=R^3r=A^2\tau+\tau^3$. 
The integral in (\ref{boundary}) diverges in the limit $\tau\to \infty$, 
but this is remedied when we subtract the boundary term
for flat space. The boundary has the topology of $S^3\times S^1$. This
is the boundary of a flat space solution where $R$ is proportional to
the distance to the origin and $r=const$. Thus, the trace of the 
extrinsic curvature is given by $3 R^{-1}$ 
and we have
\begin{equation} 
S_E^{flat}={- 3\over 8\pi G_5} \int \sqrt{\gamma} R^{-1} d^3 S^{(3)} dy.
\end{equation}
In the limit $\tau\to\infty$
we obtain
\begin{equation}
B={\pi A^2 \over 8 G}
\label{bounce}
\end{equation}
where $G = G_5/(2\pi A)$ 
is Newton's constant in four dimensions.

Thus, we find that even though flat space is metastable, the decay rate 
can be comfortably small provided that the size $A$ of the extra dimension 
is much larger than the Planck length. For instance, in the context of
M-theory, this size is of order $10^2 l_p$, and the rate (\ref{rate})
would be unobservably small, even if we multiply it by the whole volume
of our past light cone.

\section*{Singularity in four dimensions}

What is smooth in five dimensions may look singular in 
four. In this section we shall show that the solution given in the
previons section can be cast as Vilenkin's singular instanton \cite{alex}.
The singularity is of the same form as the one in Hawking and Turok's 
solution.

Compactifying the fifth dimension on a circle and using the ansatz
\begin{equation}
\tilde g_{AB}= e^{2\kappa \phi/ 3}\left(
\begin{array}{cc}
g_{\mu\nu} & 0 \\
0 & e^{-2\kappa \phi}
\end{array}
\right)\, ,
\label{ansatz}
\end{equation}
where $\kappa= (12\pi G)^{1/2}$,
the action (\ref{action5}) can be written as
\begin{equation}
S_E= \int  \sqrt{g}
\left[{1\over 2}(\partial \phi)^2 - {{\cal R}\over 16\pi G} \right]
d^4 x -{1\over 8\pi G} \int \sqrt{\gamma} K d^3\xi.
\label{action4}
\end{equation}
This coincides with the action used in Ref. \cite{alex}.

Again, we take an $O(4)\times U(1)$ symmetric ansatz for the 
metric $g_{\mu\nu}dx^{\mu}dx^{\nu}= d\sigma^2 + b^2(\sigma) dS^{(3)}$
and for $\phi=\phi(\sigma)$. With this ansatz, the field equations reduce to
\begin{equation}
\label{phieq}
\phi'' + 3{b' \over b}\phi' = 0
\end{equation}
\begin{equation}
\left({b'\over b}\right)^2=  {4\pi G \over 3}\phi'^2 + {1\over b^2},
\label{beq}
\end{equation}
where primes stand for derivatives with respect to $\sigma$.
From the four dimensional point of view, the instanton (\ref{instanton})
corresponds to 
\begin{equation}
\label{sol}
b = \tau^{1/2} (A^2+\tau^2)^{1/4}
\end{equation}
\begin{equation}
\phi  = {-3\over 4\kappa} \ln\left({\tau^2 \over A^2+\tau^2}\right).
\label{solphi}
\end{equation}
Using
$|{d\tau/d\sigma}| =  (A^2+\tau^2)^{1/4}\tau^{-1/2}$,
it is straightforward to check that (\ref{sol}-\ref{solphi})
satisfy the field equations (\ref{phieq}-\ref{beq}).  

At large $\tau$, we have  $\sigma \approx \tau$. Therefore
$b\approx \sigma$ and  $\phi\approx C/(2 \sigma^{2})$, where
\begin{equation}
C=3 A^2 / 2\kappa, \label{c}
\end{equation} 
Near $\tau=0$ we have 
$\tau^{3/2} \approx (3A^{1/2}/2) (\sigma-\sigma_f)$, where 
$\sigma_f$ is a constant. Substituting in
(\ref{sol}) we find that
$\phi\approx -\kappa^{-1} \ln (\sigma-\sigma_f)+ const.$, and 
$b^3\approx C \kappa (\sigma-\sigma_f)$. Therefore, our regular
instanton looks exactly like Vilenkin's singular solution
when viewed in four dimensions.

\section*{Singular instantons and the boundary term}

As mentioned above, the instanton (\ref{instanton}) looks
singular at $\sigma=\sigma_f$ when viewed in four dimensions.
It was generally believed \cite{alex,HT,all} that the contribution of 
the singularity to the Euclidean action should be given by the 
Hawking-Gibbons boundary term [i.e., the last term in (\ref{action4})] 
evaluated at the singular boundary. For the solution given above, this
action is given by \cite{alex}
\begin{equation}
S_{GH}= 3 \pi^2 \kappa^{-1} C= {3\pi A^2 \over 8 G} 
\label{boundary2}
\end{equation}
However, this procedure is not very well motivated within the 
framework of the ``no-boundary'' proposal, where the boundary is not 
supposed to be there in the first place. Indeed,
the naive result (\ref{boundary2}) is off by a factor 
of 3 of the correct result given in (\ref{bounce}).

In ref. \cite{ga98} we suggested that, like most time-like singularities, 
the one in Hawking and Turok's solution should be regularized
with matter. In this way, the singular instanton can be viewed as the limit
of a family of no boundary solutions where both the scalar field and the
geometry are regular. When this is done, it is found that the contribution 
of the singularity to the Euclidean action is given by 1/3 of the 
Gibbons-Hawking term at the singular boundary, in agreement with the
result (\ref{bounce}).

One may worry that boundary contributions in four and five 
dimensions do not add up to be the same. This is not the case. The difference
between (\ref{bounce}) and (\ref{boundary2}) is due to the fact that in
four dimensions a new boundary was added at $\tau=0$. It can be
easily checked that if the point $\tau=0$ is removed from the smooth 
manifold described by (\ref{instanton}), there is a new
boundary term which accounts for the difference.
However, such ad-hoc operation would be unjustifiable. If the point $\tau=0$ 
is excised, nothing links the parameter $A$ with the asymptotic radius of the 
extra dimension. As a result, the parameter can be chosen as small as desired
and the Euclidean action can be made arbitrarily small, which would mean that
the decay of flat spacetime would be unsuppressed. 

\section*{Open inflation}

In order to find inflationary solutions, a potential must be added
to the action (\ref{action4}). In the context of Kaluza-Klein theories,
it is believed that the dilaton will be stabilized by an
effective potential generated by quantum corrections. It is possible that
this same potential may drive inflation in the appropriate range. 
Here, we shall not attempt to enter
into a detailed discussion of this possibility. Instead, we note that
a cosmological constant in the five dimensional theory acts 
as an exponential potential for $\phi$ in the four dimensional theory.
Denoting by $\Lambda$ the cosmological constant in five dimensions, 
the potential in four dimensions is given by
\begin{equation}
V(\phi)=\lambda e^{2\kappa \phi/3}. \label{expo}
\end{equation}
Here, $\lambda= 2\pi A \Lambda$, where $A$ is the coordinate 
radius of the
extra dimension. The rest of the four-dimensional action (\ref{action4}) 
remains unchanged by this addition. 
A potential of the form (\ref{expo})
drives power-law inflation \cite{luma}, with power exponent $p=3$ \cite{mo}. 

Let us now turn to the solution of the model. This is straightforward in
five dimensions. With the ansatz (\ref{fivemetric}) the equations of
motion reduce to $\ddot X + \alpha X = 2k,$
and $r\propto \dot R$, where $X\equiv R^2$, $\alpha\equiv 2\Lambda/3$, 
$k=1$ for the three-sphere and the dot
indicates derivative with respect to $\tau$. The only solution which is
regular everywhere is given by
\begin{equation}
\label{halfsphere}
R=2 \alpha^{-1/2}\sin(\alpha^{1/2} \tau/2)
\end{equation}
$$
r=2\alpha^{-1/2}A^{-1} \cos(\alpha^{1/2} \tau/2).
$$
The parameter $A$ is irrelevant: it is elliminated by changing the
fifth coordinate to an angular coordinate $\theta_5=y/A$ which runs
from $0$ to $2\pi$.
The solution (\ref{halfsphere}) has an interesting topology. 
Near $\tau=0$ the manifold looks like the product of an open set of
$R^4$ with a circle of radius $\hat A\equiv 2 \alpha^{-1/2}$. 
However, as $\tau$
progresses, the radius of the circle shrinks, and finally closes
smoothly at $\tau=\hat A\pi/2$. There, the manifold looks like the product
of an open set of $R^2$ whith the three-sphere of radius 
$\hat A$. Hence, it appears that every closed path and every closed 
three-surface on the manifold can be deformed to a point by bringing
them to the appropriate end of the $\tau$ range. 
The subspace $y=const$ is half of a four-sphere, whereas
the subspace spanned by the coordinates $\tau$ and $y$ while keeping
the rest fixed is half of a two-sphere. 

Let us now look at our instanton from the point of view of the 
four-dimensional ansatz. Near $\tau=0$ we have 
$\sigma\approx \tau$, $\phi\approx (3\sigma^2/4\kappa \hat A)$ and
$b\approx \sigma$. Therefore, the solution is perfectly regular from
the four-dimensional point of view too. However, near $\tau=\hat A\pi/2$
we have $\tau-(\hat A \pi/2) \propto (\sigma_f - \sigma)^{2/3}$,
where $\sigma_f=\hat A(2/\pi)^{1/2}\Gamma^2(3/4)$ is the value of the 
coordinate $\sigma$ at $\tau=\hat A\pi/2$, and 
$b^3 \approx \hat C(\sigma_f-\sigma)$,
$\phi\approx -\kappa^{-1} \ln(\sigma_f-\sigma)+const$. Here 
$\hat C=3\hat A^2/ 2\kappa$. Therefore, the singularity is of the same
form as the one in Hawking and Turok's model. Notice that the four dimensional
``scale factor'' $b(\sigma)$ goes to zero at $\sigma_f$ too.
The evolution of the universe after nucleation is given by the analytic
continuation of our instanton through the complexification 
$\psi\to (\pi/2) - i \hat \psi$.
The resulting chart covers all of the spacetime,
except for the the future light cone from the center of symmetry $\sigma=0$.
The open chart covering this region is obtained by taking 
$\sigma \to i \hat\sigma$ and $\psi\to -i \hat\psi$, and it represents the
inflating open universe from where the observable universe would emerge.

\section*{Conclusions}

We have constructed a smooth five-dimensional instanton
representing the creation of an open universe. When reduced to four
dimensions, the solution looks like a singular instanton of the type
introduced by Hawking and Turok. Although our construction seems to
validate the use of such instantons, some words of caution must be
said.

The smooth instanton presented above exists only when
the value of inflaton at $\sigma=0$ is $\phi=0$.
Otherwise, there will be a conical singularity at the 
point where the fifth dimension closes. The same would happen for a more 
general potential: the fifth dimension would only close smoothly for a 
particular value of $\phi$ at $\sigma=0$. Hence, it appears that these
instantons cannot produce a range of values of the density parameter $\Omega$. 
This casts some doubt on the method used in \cite{HT} to find the
probability distribution for $\Omega$. On the positive side, it is still
true that one can obtain an open universe without requiring a special form 
of the inflaton potential.
We should add that there are more ``conventional'' ways of obtaining a range of
values of $\Omega$ in theories where one field undergoes a first order
phase transition and a second field is responsible for slow roll inflation
inside the nucleated bubbles \cite{LM}. In such models, one finds that 
a range of values of $\Omega$ occurs inside of {\em each} nucleated bubble 
\cite{GGM}, and depending on parameters of the model, it is not 
unlikely for an observer to measure the density parameter in the range 
$(1-\Omega)/\Omega \sim 1$ \cite{GTV}. 

We have also found that flat space with a compact extra dimension is
gravitationally metastable. It decays through nucleation of bubbles
of ``nothing'' which eat up spacetime as they expand. The nucleation
rate is unobservably small provided that the size of the extra dimension
is large compared with the Planck length.

It is a pleasure to thank Xavier Montes, Alex Pomarol, Takahiro Tanaka,
Enric Verdaguer and Alex Vilenkin for very useful conversations. The 
author acknowledges support from CICYT under contract AEN95-0882 and from 
NATO under grant CRG 951301.


\begin{thebibliography}{999}
\bibitem{mo} Power law open inflation has some appeal from the 
phenomenological point of view, since it gives a tilted spectrum of 
density perturbations \cite{blue}. As it stands, however, the model 
(\ref{expo}) is unrealistic because reheating would never occur. 
Here, shall not be too concerned about this fact. The model will be 
used only as an example where the equations can be solved 
analytically.
\end{thebibliography}
\end{document}